\documentclass[preprint,review,12pt]{elsarticle}

\usepackage{graphics}

\usepackage{amssymb}

\usepackage{lineno}

\journal{Icarus}

\begin{document}

\begin{frontmatter}

\title{Equilibrium composition between liquid and clathrate reservoirs on Titan}

\author{Olivier~Mousis}
\ead{olivier.mousis@obs-besancon.fr}
\address{Universit{\'e} de Franche-Comt{\'e}, Institut UTINAM, CNRS/INSU, UMR 6213, Besan\c con Cedex, France}
\address{Center for Radiophysics and Space Research, Space Sciences Building, Cornell University,  Ithaca, NY 14853, USA}
\author{Mathieu Choukroun}
\address{Jet Propulsion Laboratory, California Institute of Technology, Pasadena, California, USA}
\author{Jonathan I. Lunine}
\address{Center for Radiophysics and Space Research, Space Sciences Building, Cornell University,  Ithaca, NY 14853, USA}
\author{Christophe Sotin}
\address{Jet Propulsion Laboratory, California Institute of Technology, Pasadena, California, USA}

\begin{abstract}
Hundreds of lakes and a few seas of liquid hydrocarbons have been observed by the Cassini spacecraft to cover the polar regions of Titan. A significant fraction of these lakes or seas could possibly be interconnected with subsurface liquid reservoirs of alkanes. In this paper, we investigate the interplay that would happen between a reservoir of liquid hydrocarbons located in Titan's subsurface and a hypothetical clathrate reservoir that progressively forms if the liquid mixture diffuses throughout a preexisting porous icy layer. To do so, we use a statistical--thermodynamic model in order to compute the composition of the clathrate reservoir that forms as a result of the progressive entrapping of the liquid mixture. This study shows that clathrate formation strongly fractionates the molecules between the liquid and the solid phases. Depending on whether the structure I or structure II clathrate forms, the present model predicts that the liquid reservoirs would be mainly composed of either propane or ethane, respectively. The other molecules present in the liquid are trapped in clathrates. Any river or lake emanating from subsurface liquid reservoirs that significantly interacted with clathrate reservoirs should present such composition. On the other hand, lakes and rivers sourced by precipitation should contain higher fractions of methane and nitrogen, as well as minor traces of argon and carbon monoxide.

\end{abstract}

\begin{keyword}
Titan -- Titan, hydrology -- Titan, surface -- Titan, atmosphere
\end{keyword}

\end{frontmatter}


\section{Introduction}
\label{intro}

In 1811, Sir Humphrey Davy (Davy 1811) was the first to report the existence of clathrate, a variety of compounds in which water forms a continuous and known crystal structure with small cages. These cages trap guests, such as methane or ethane, needed to stabilize the water lattice. The two most common clathrate structures found in nature are known as structures I and II, which differ in the type of water cages present in the crystal lattice (Sloan \& Koh 2008). Structure I has two types of cages, a small pentagonal dodecahedral cage, denoted 5$^{12}$ (12 pentagonal faces in the cage) and a large tetrakaidecahedral cage, denoted 5$^{12}$6$^2$ (12 pentagonal faces and 2 hexagonal faces in the cage). Structure II also has two types of cages, a small 5$^{12}$ cage and a large hexakaidecahedral cage, denoted 5$^{12}$6$^4$ (12 pentagonal faces and 4 hexagonal faces in the cage). The type of structure that forms depends largely on the size of the guest molecule. For example, methane and ethane induce water to form structure I clathrate and propane structure II clathrate (Sloan \& Koh 2008).

On Titan, the temperature and atmospheric pressure conditions prevailing at the ground level permit clathrates formation when liquid hydrocarbons enter in contact with the exposed water ice (Mousis \& Schmitt 2008). Assuming an open porosity for Titan's upper crust, clathrates made from hydrocarbons are even expected to be stable down to several kilometers from the surface (Mousis \& Schmitt 2008). An interesting feature of clathrates is that their formation induces a fractionation of the trapped molecules (van der Waals \& Platteuw 1959; Lunine \& Stevenson 1985; Mousis et al. 2010). This property has been used to suggest that the noble gas depletion observed in Titan's atmosphere could result from their efficient sequestration in a global clathrate layer located in the near subsurface (Thomas et al. 2007; Mousis et al. 2011). It has also recently been used to propose that the satellite's polar radius, which is smaller by several hundred meters than the value predicted by the flattening due to its spin rate, would result from the substitution of methane by ethane percolating in liquid form in clathrate layers potentially existing in the polar regions (Choukroun \& Sotin 2012).

In this paper, we investigate the interplay that may happen between an alkanofer, namely a reservoir of liquid hydrocarbons located in Titan's subsurface, and a hypothetical clathrate reservoir that progressively forms when the liquid mixture diffuses throughout a preexisting porous icy layer. This porous layer might have been generated by cryovolcanic events resulting from the ascent of liquid from subsurface ocean (Mitri et al. 2006) or from the destabilization of clathrates in the ice shell of Titan (Tobie et al. 2006). In both cases, a highly porous icy material in contact with the atmosphere is generated, probably similar to basaltic lava flows (Mousis \& Schmitt 2008). The cooling of cryolava is expected to take less than 1 yr to decrease down to Titan's surface temperature (Lorenz 1996), implying that it should be fast enough to allow preservation of the created porosity.

Hundreds of lakes and a few seas are observed to cover the polar regions of Titan (Stofan et al. 2007). Kraken Mare and Ligeia Mare, namely the two largest seas of Titan, have surface areas estimated to be $\sim$400,000 km$^2$ (Jaumann et al. 2010) and 126,000 km$^2$ (Mastrogiuseppe et al. 2014), respectively. With an average depth of $\sim$70 m, Ligeia Mare contains $\sim$5 $\times$ 10$^{15}$ kg of hydrocarbons, about 100 times the known terrestrial oil and gas reserves, but still only $\sim$1.4\% of Titan's atmospheric methane (Mastrogiuseppe et al. 2014). While a significant number of these lakes and seas should be regularly filled by hydrocarbon rainfalls (Turtle et al. 2011), some of them could be also renewed via their interconnection with alkanofers. Using porous media properties inferred from Huygens probe observations, Hayes et al. (2008) found that the timescales for flow into and out of observed lakes via subsurface transport are order of tens of years. Because the porosity is not expected to evolve significantly over $\sim$20 Myr within the subsurface of Titan (Kossacki \& Lorenz 1996), clathrates might form and equilibrate with liquid hydrocarbons well prior that the porosity reaches its close-off value. A fraction of these lakes may then result from the interaction between alkanofers and clathrate reservoirs through the ice porosity and possess a composition differing from that of lakes and rivers sourced by precipitation. As a liquid reservoir occupies a finite volume, the progressive transfer and fractionation of the molecules in the forming clathrate reservoir could alter the lakes' chemical composition. 

In order to explore this possibility, we use a statistical--thermodynamic model derived from the approach of van der Waals \& Platteuw (1959) to compute the composition of the clathrate reservoir that forms as a result of the progressive entrapping of the species present in the liquid mixture. The major ingredient of our model is the description of the guest--clathrate interaction by a spherically averaged Kihara potential with a set of potential parameters based on the literature. This allows us to track the evolution of the mole fractions of species present in the liquid reservoir as a function of their progressive entrapment in the clathrate layer. Section 2 is devoted to the description of our computational approach and the physical ingredients of our model. We also discuss the underlying assumptions of our approach in this Section. The results concerning the composition of lakes interacting with clathrate reservoirs at polar or equatorial zones are presented in Section 3. Section 4 is devoted to discussion and conclusions.

\section{Modeling the clathrate--liquid equilibrium}
\label{mode}

\subsection{Computational approach}

\begin{figure}
\begin{center}
\resizebox{\hsize}{!}{\includegraphics[angle=0]{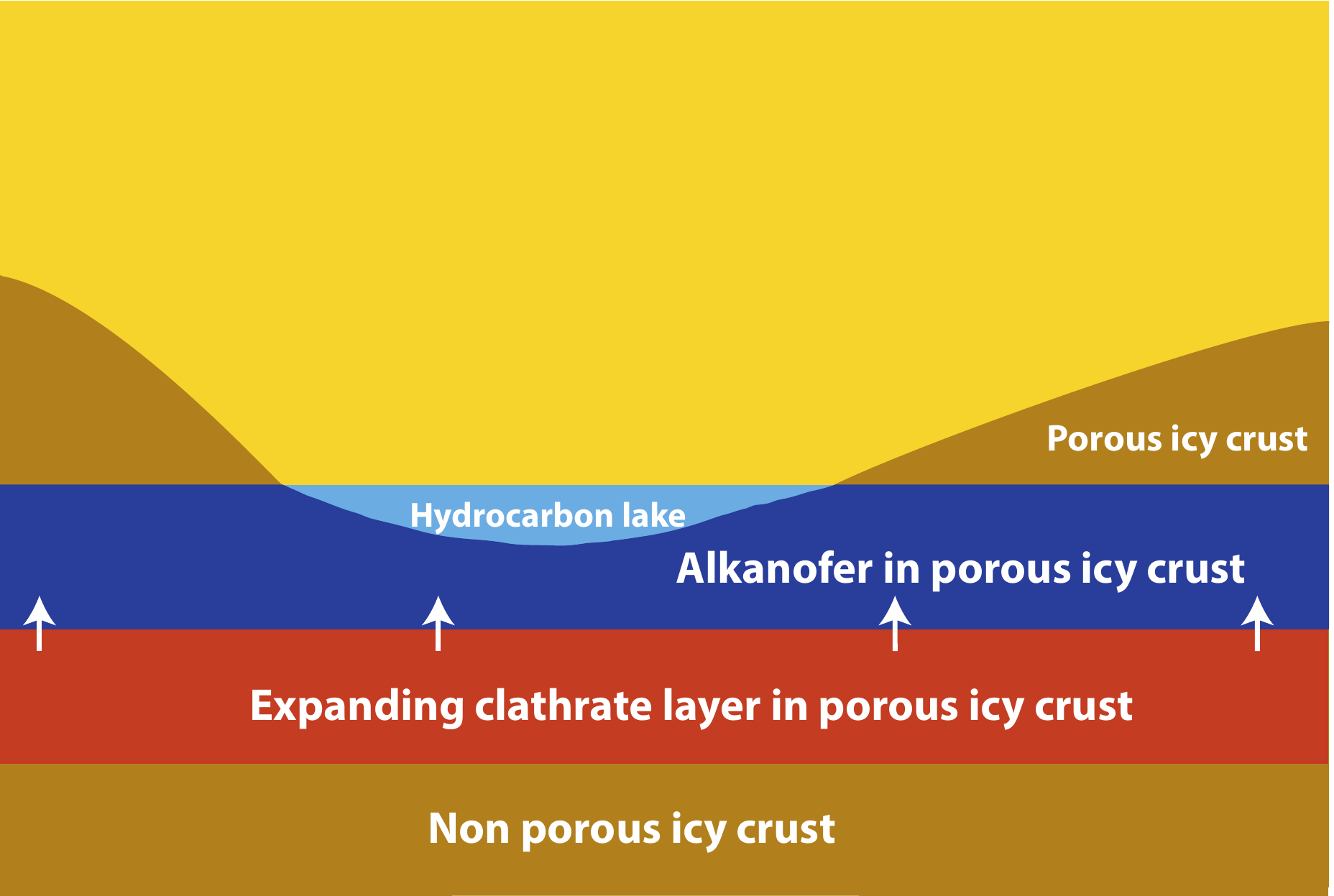}}
\caption{Schematic view of the model. The alkanofer and clathrate reservoir are assumed to communicate and form an isolated system that occupies the open porosity of the icy crust. The clathrate reservoir expands with time throughout the porous icy crust at the expense of the alkanofer. The progressive transfer and fractionation of the molecules in the forming clathrate reservoir alter the alkanofer's chemical composition. In some circumstances (see Sec. Discussion), the lake's composition might reflect that of the alkanofer from which it is sourced.}
\label{Draw}
\end{center}
\end{figure}

We assume that the liquid reservoir is in contact with porous ice and that clathrates form at the liquid/ice interface. We consider an isolated system composed of a clathrate reservoir that progressively forms and replaces the H$_2$O crustal material with time and a liquid reservoir that correspondingly empties due to the net transfer of molecules to the clathrate reservoir (see concept pictured in Fig. \ref{Draw}). Based on this approach, we have elaborated a computational procedure with the intent to determine the mole fractions of each species present in the liquid reservoir and trapped in the forming clathrate reservoir, as a function of the fractions of the initial liquid volume (before volatile migration) remaining in lake and present in clathrates, respectively. At the beginning of our computations, the liquid reservoir's composition is derived from those computed by Cordier et al. (2009, 2013) for lakes at polar and equator temperatures, which result from models assuming thermodynamic equilibrium between the atmosphere and the lakes (see Table \ref{lake0}). 

Because the clathration kinetics of hydrocarbons is poorly constrained at the Titan's surface temperatures considered, the present calculations use an iterative process for which the number of molecules in the liquid phase being trapped in clathrates between each iteration is equal to 10$^{-4}$ the total number of moles. Initially, all molecules are in the liquid phase. The mole fraction $x^{\rm lake}_K (0)$ of species $K$ in lake, is given in Table \ref{lake0}. The corresponding number of moles of species $K$ is defined by $n^{\rm lake}_K (0)$~=~$x^{\rm lake}_K (0)$~$\times$~$n_{\rm lake} (0)$, with $n_{\rm lake} (0)$ the number of moles of liquid available in the lake at this time. The mole fraction $x_{K}^{\rm{clat}}(0)$ of the enclathrated species $K$ and the number of moles $n_{\rm clat} (0)$ of liquid trapped in clathrate are set to zero.

At iteration $i$, the mole fraction $x_{K}^{\rm{clat}}(i)$ of each enclathrated guest $K$ is calculated by using the statistical-thermodynamic model described in section \ref{stat} and the relative abundances in the liquid phase of the previous iteration. The new numbers of moles in the lake $n^{\rm lake}_K (i)$ and in clathrate $n^{\rm clat}_K (i)$ are calculated for each species $K$, with $n^{\rm lake}_K (i)$~=~$n^{\rm lake}_K (i~-~1)$~-~$x^{\rm clat}_K (i)$~$\times$~10$^{-4}$ and $n^{\rm clat}_K (i)$~=~$n^{\rm clat}_K (i~-~1)$~+~$x^{\rm clat}_K (i)$~$\times$~10$^{-4}$. The mole fraction of each species $K$ present at iteration $i$ in the lake and clathrate are defined by $x_{K}^{\rm{lake}}(i)~=~\frac{n_{K}^{\rm{lake}}(i)}{n_{\rm lake}(i)}$ and $f_{K}^{\rm{clat}}(i)~=~\frac{n_{K}^{\rm{clat}}(i)}{n_{\rm clat}(i)}$, respectively, with $n_{\rm lake}$($i$)~=~$\displaystyle\sum_K$ $n^{\rm lake}_K$($i$) and $n_{\rm clat}$($i$)~=~$\displaystyle\sum_K$~ $n^{\rm clat}_K$($i$). At any iteration $i$, $n_{\rm tot}$ = $n_{\rm lake}$($i$) + $n_{\rm clat}$($i$). The new values of $n^{\rm lake}_K (t)$ and $n^{\rm clat}_K (t)$ are introduced in the next loop and the process is run until $n_{\rm lake}$ eventually gets to zero.

\subsection{The statistical-thermodynamic model}
\label{stat}

To calculate the relative abundances of guest species incorporated in the clathrate phase at given temperature and pressure, we use a model applying classical statistical mechanics that relates the macroscopic thermodynamic properties of clathrates to the molecular structure and interaction energies (van der Waals \& Platteuw 1959; Lunine \& Stevenson 1985; Mousis et al. 2010). It is based on the original ideas of van der Waals and Platteeuw for clathrate formation, which assume that trapping of guest molecules into cages corresponds to the three-dimensional generalization of ideal localized adsorption (see Sloan \& Koh (2008) for an exhaustive description of the statistical thermodynamics of clathrate equilibra). 
	
In this formalism, the fractional occupancy of a guest molecule $K$ for a given type $q$ ($q$~=~small or large) of cage can be written as

\begin{equation}
y_{K,q}=\frac{C_{K,q}f_K}{1+\sum_{J}C_{J,q}f_J} ,
\label{eq1}
\end{equation}

\noindent where the sum in the denominator includes all the species which are present in the liquid phase. $C_{K,q}$ is the Langmuir constant of species $K$ in the cage of type $q$, and $f_K$  the fugacity of species $K$ in the mixture. Using the Redlich-Kwong equation of state (Redlich and Kwong 1949) in the case of a mixture dominated by C$_2$H$_6$, we find that the coefficient of fugacity $\phi$ of the mixture (defined as the ratio of the mixture's fugacity to its vapor pressure) converges towards 1 at Titan's surface temperatures and corresponding C$_2$H$_6$ vapor pressures. In our approach, the value $f_K$ of each species $K$ is calculated via the Raoult's law, which states 

\begin{equation}
f_K \simeq  P_K = x^{lake}_K \times P^*_K,
\label{Raoult}
\end{equation}

\noindent with $P_K$ the vapor pressure of species $K$ in the mixture and $P^*_K$  the vapor pressure of pure component $K$. $P^*_K$ is defined via the Antoine equation

\begin{equation}
log_{\rm 10}(P^*_K)~=~{\rm A}~-~\frac{{\rm B}}{T~+~ {\rm C}},
\label{Ant}
\end{equation}

\noindent with the parameters A, B and C listed in Table \ref{ABC} ($P^*_K$ is expressed in bar and $T$ in K).

The Langmuir constant depends on the strength of the interaction between each guest species and each type of cage, and can be determined by integrating the molecular potential within the cavity as

\begin{equation}
C_{K,q}=\frac{4\pi}{k_B T}\int_{0}^{R_c}\exp\Big(-\frac{w_{K,q}(r)}{k_B T}\Big)r^2dr ,
\label{eq2}
\end{equation}

\noindent where $R_c$ represents the radius of the cavity assumed to be spherical, $k_B$ the Boltzmann constant, and $w_{K,q}(r)$ is the spherically averaged Kihara potential representing the interactions between the guest molecules $K$ and the H$_2$O molecules forming the surrounding cage $q$. This potential $w(r)$ can be written for a spherical guest molecule, as (McKoy \& Sinano\u{g}lu 1963)

\begin{eqnarray}
w(r) 	= 2z\epsilon\Big[\frac{\sigma^{12}}{R_c^{11}r}\Big(\delta^{10}(r)+\frac{a}{R_c}\delta^{11}(r)\Big) \\ \nonumber
- \frac{\sigma^6}{R_c^5r}\Big(\delta^4(r)+\frac{a}{R_c}\delta^5(r)\Big)\Big],
\label{eq3}
\end{eqnarray}

\noindent with

\begin{equation}
\delta^N(r)=\frac{1}{N}\Big[\Big(1-\frac{r}{R_c}-\frac{a}{R_c}\Big)^{-N}-\Big(1+\frac{r}{R_c}-\frac{a}{R_c}\Big)^{-N}\Big].
\label{eq4}
\end{equation}

\noindent In Eq. 5, $z$ is the coordination number of the cell. This parameter depends on the structure of the clathrate (I or II;  see Sloan \& Koh 2008) and on the type of the cage (small or large). The Kihara parameters $a$, $\sigma$ and $\epsilon$ for the molecule-water interactions, given in Table \ref{Kihara}, have been taken from the recent compilation of Sloan \& Koh (2008) when available and from Parrish \& Prausnitz (1972) for the remaining species.

Finally, the mole fraction $x^{clat}_K$ of a guest molecule $K$ in a clathrate can be calculated with respect to the whole set of species considered in the system as

\begin{equation}
x^{clat}_K=\frac{b_s y_{K,s}+b_\ell y_{K,\ell}}{b_s \sum_J{y_{J,s}}+b_\ell \sum_J{y_{J,\ell}}},
\label{eq5}
\end{equation}

\noindent where $b_s$ and $b_l$ are the number of small and large cages per unit cell respectively, for the clathrate structure under consideration, and with ${\sum_{K}} f_{K}~=~1$. Values of $R_c$, $z$, $b_s$ and $b_l$ are taken from Parrish \& Prausnitz (1972). 

Among the different species considered in the present study, C$_3$H$_8$ is the only molecule whose size is too large to be trapped either in small or large cages of structure I clathrate. Because C$_3$H$_8$ can only be trapped in the large cages of structure II clathrate (Sloan \& Koh 2008), we assume that this molecule remains in the liquid phase in the case of structure I clathrate formation.

$\sigma_{K-W}$ is the Lennard-Jones diameter, $\epsilon_{K-W}$ is the depth of the potential well, and $a_{K-W}$ is the radius of the impenetrable core, for the guest-water pairs.

\subsection{Model uncertainties}
\label{uncert}

The scenario we propose is based on some underlying assumptions, indicated below:

\begin{itemize}

\item {\it Statistical thermodynamic model}. The predictive capabilities of our model, which is derived from the approach of van der Waals \& Platteeuw, rely on four key assumptions: (i) the host molecules contribution to the free energy is independent of the clathrate occupancy (the guest species do not distort the cages); (ii) the cages are singly occupied; (iii) guest molecules rotate freely within the cage and they do not interact with each other; (iv) classical statistics is valid, i.e., quantum effects are negligible. However, these assumptions are subject to caution since encaged molecules with larger dimensions may distort the cages. Also, for certain small-sized molecules (like H$_2$) multiple gas occupancy can occur, and non spherical molecules may not be free to rotate in the entire cavity. Molecular dynamics simulations (Erfan-Niya et al. 2011; Fleischer \& Genda 2013) are typically used to investigate these effects but, due to the amount of time they require, these computations do not easily provide quantitative estimates on the fractionation of the different species encaged in clathrates at the macroscopic level, in particular in systems considering a large number of species. For these reasons, and because it is often based on interaction parameters fitted on laboratory measurements of phase equilibria, allowing  accurate prediction when compared to experiments, the approach of van der Waals \& Platteeuw remains the main tool employed in industry and research to determine clathrate composition (Sloan \& Koh 2008).

\item {\it Kinetics of clathrate formation.} Kinetics data concerning clathrate formation are scarce and mostly concern gas/ice interaction (see Sloan \& Koh for a review of measurements). In the present case, clathrates form from the interaction between liquid hydrocarbons and ice. To the best of our knowledge, the kinetics measurements of the closest system reported so far are those concerning clathrate formation from a mixture of liquid methane and ethane at temperatures ranging from 260 to 280 K (Murshed et al. 2010). Because of the large temperature difference between Titan's surface and these experiments, the uncertainty is too large to make use of these kinetics data. Kinetics measurements remain to be done at Titan's conditions.

\item {\it Assumption of equilibrium.} Our model is restricted to equilibrium calculations between subsurface alkanofers and coexisting clathrate layers. Hence it should be applied with caution to the case where lakes and rivers located on Titan's surface directly equilibrate with a clathrate layer located beneath. To do so, we would need to compute the simultaneous equilibrium between the clathrate reservoir, lake and atmosphere. However, our computations are a good approximation if the reequilibration timescale between the lake and the atmosphere is short compared to the timescale of clathrate formation.

\end{itemize}

\section{Results}

\begin{figure*}
\begin{center}
\resizebox{\hsize}{!}{\includegraphics[angle=0]{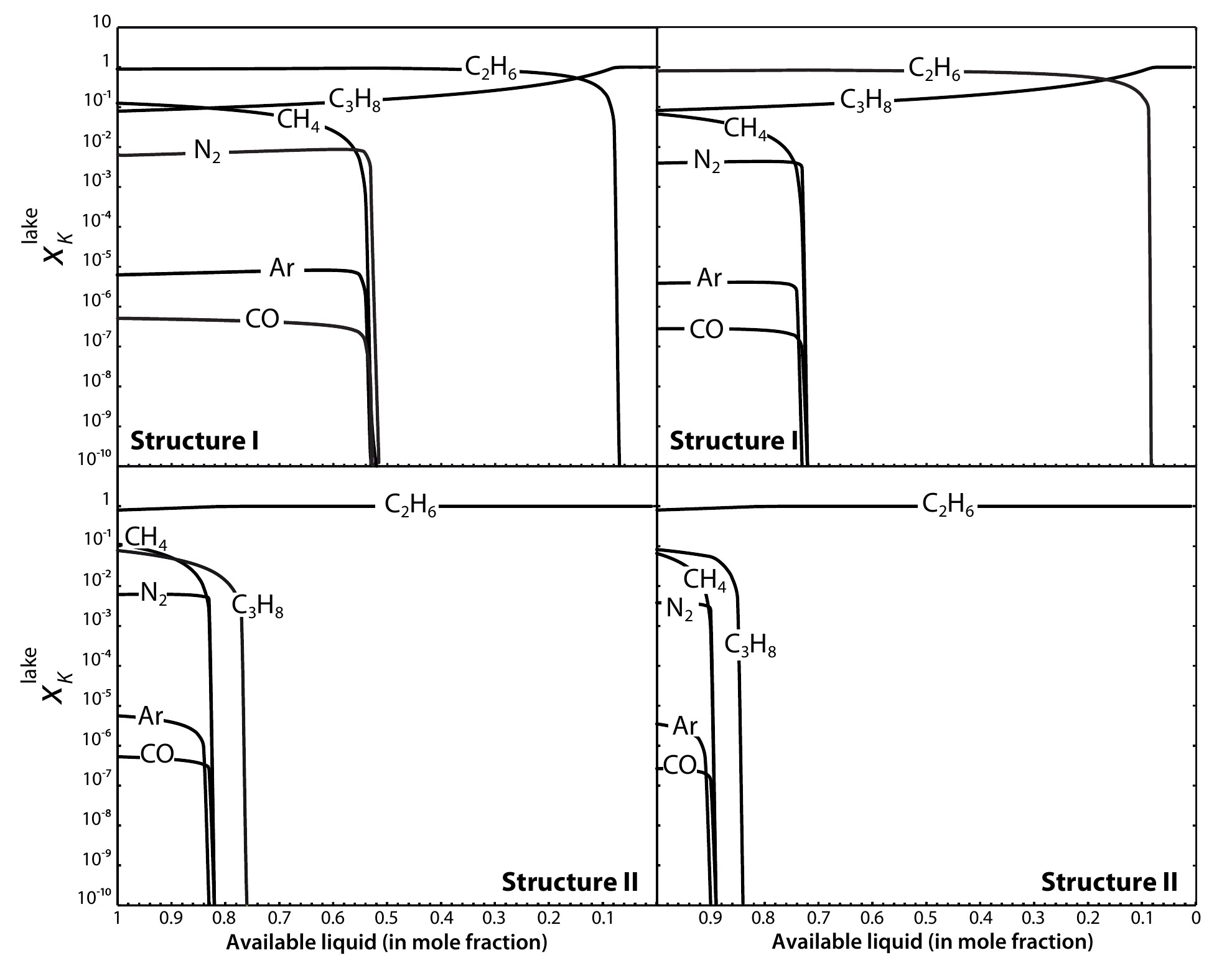}}
\caption{Composition of alkanofers located at the pole (left panels) and at the equator (right panels) in contact with forming clathrate layers. Mole fractions $x_{K}^{\rm{lake}}$ of various species $K$ in solution depend on the mole fraction of available liquid (number of moles still present in the liquid phase to the total number of moles present in the initial liquid reservoir). Top and bottom panels correspond to the formation of structure I and II clathrates, respectively.}
\label{lake}
\end{center}
\end{figure*}

Figure \ref{lake} represents the evolution of the mole fractions of species present in subsurface alkanofers of Titan, starting with the one given in Table \ref{lake0}, as a function of their progressive entrapping in structures I and II clathrate reservoirs located at the poles ($T$ $\sim$90 K) and at the equator ($T$ $\sim$93.6 K). The evolution of the liquid reservoir's composition varies significantly if one assumes the formation of a structure I or a structure II clathrate reservoir. In particular, the change of clathrate structure in our model alters the number of dominant species present in the liquid phase at high mole fractions of entrapped liquid. It also drastically affects the evolution of the abundances of secondary species during the progressive liquid entrapping.

When considering the formation of a structure I clathrate, and irrespective of the liquid reservoir's temperature, the dominating species is C$_2$H$_6$ until that a liquid mole fraction of $\sim$0.85 has been trapped into clathrate. Above this value, and because of the entrapping of the other molecules in structure I clathrate, C$_3$H$_8$ becomes the only remaining species in the liquid reservoir. At both temperatures considered, the initial abundance of CH$_4$ in the liquid phase is close to that of C$_3$H$_8$ (see Table \ref{lake0}). However, as the liquid progressively forms clathrate with ice, the mole fraction of CH$_4$ rapidly decreases and finally converges towards zero after a mole fraction of $\sim$0.3--0.5 of the initial liquid reservoir has been entrapped. Meanwhile, the mole fractions of N$_2$, Ar and CO form plateaus and finally drop towards zero at the same mole fraction of entrapped liquid that corresponds to the disappearance of CH$_4$.
 
When considering the formation of a structure II clathrate, the dominant species remains C$_2$H$_6$, irrespective of the mole fraction of liquid entrapped in clathrate and the temperature of the reservoirs. Instead of increasing with the progressive liquid entrapment in clathrate as in the previous case, the abundance of C$_3$H$_8$ decreases and suddenly drops when the mole fraction of entrapped liquid is $\sim$0.15--0.23. The abundances of CH$_4$, N$_2$, Ar and CO also decrease with the progressive formation of structure II clathrate and suddenly drop at mole fractions of entrapped liquid in the $\sim$0.11--0.18 range. The temperature of the liquid and clathrate reservoirs also plays a role in the determination of their composition, but in a less important manner than the modification of the clathrate structure. Figure \ref{lake} shows that the rise of temperature decreases by several per cents the mole fraction of entrapped liquid at which the abundances of minor species drop in the solution. Compared to the change of clathrate structure, a temperature variation also affects (to a lower extent) the mole fractions of secondary species in the liquid reservoir but does not influence those of major compounds.

\begin{figure*}
\begin{center}
\resizebox{\hsize}{!}{\includegraphics[angle=0]{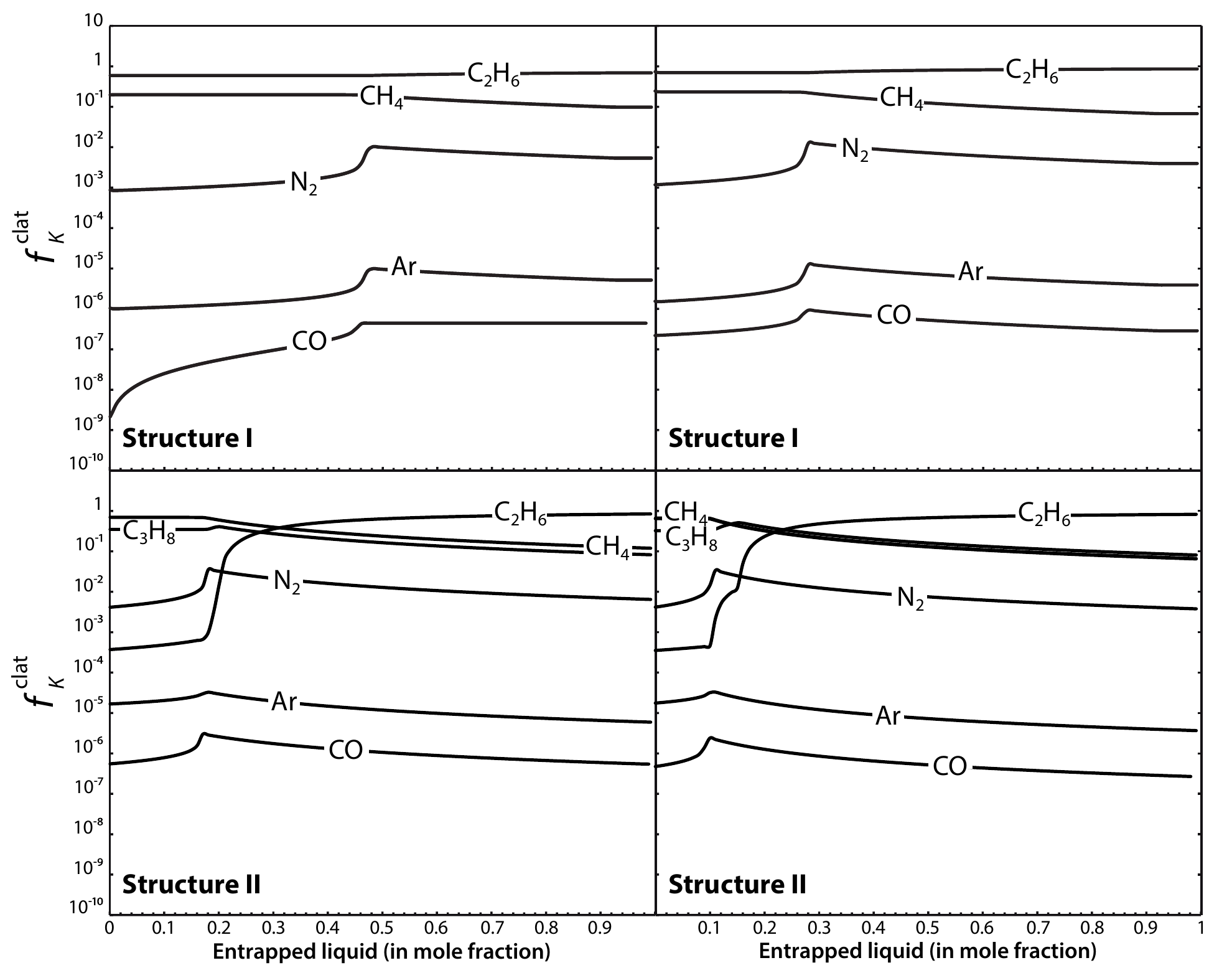}}
\caption{Composition of clathrate layers forming in contact with alkanofers located at the pole (left panels) and at the equator (right panels). Mole fractions $f_{K}^{\rm{clat}}$ of various species $K$ trapped in clathrate depend on the mole fraction of entrapped liquid (number of moles entrapped in clathrate to the total number of moles present in the initial liquid reservoir). Top and bottom panels correspond to the formation of structure I and II clathrates, respectively. C$_3$H$_8$ is only trapped in structure II clathrate (see text).}
\label{clat}
\end{center}
\end{figure*}

Figure \ref{clat} represents the evolution of the composition of structures I and II clathrate reservoirs on Titan as a function of the progressive entrapping of subsurface alkanofers located at the poles and at the equator. As noted for the composition of the liquid reservoirs, the structure change affects the number of dominant species in clathrate. It also significantly affects the evolution of the mole fractions of secondary species during their progressive trapping. The mole fractions of the different entrapped species strongly differ from those in solution at the beginning of their entrapping, as a result of the fractionation occurring during the clathration. However, irrespective of the structure considered and because of the conservation of the total number of moles in our system, the mole fraction of each species trapped in clathrate converges towards its initial abundance in solution when the fraction of entrapped liquid approaches 1.

In the case of structure I clathrate formation, C$_2$H$_6$ remains the dominant volatile. On the other hand, with a mole fraction ranging between 0.11 and 0.25, CH$_4$ is the second most abundant volatile present in clathrate, except at fractions of entrapped liquid close to 1 and at equator temperatures. With a mole fraction in the 10$^{-3}$--10$^{-2}$ range, N$_2$ is the third most abundant volatile trapped in clathrate at fractions of entrapped liquid lower than $\sim$0.9. The mole fractions of Ar and CO are in the 10$^{-6}$--10$^{-5}$ and 10$^{-9}$--10$^{-6}$ ranges, respectively, making them the less abundant species present in the forming structure I clathrate. As mentioned in Sec. \ref{mode}, C$_3$H$_8$ is not trapped in structure I clathrate, due to its large size compared to those of small and large cages.

When considering the formation of structure II clathrate, the two most abundant species present in clathrate are CH$_4$ and C$_3$H$_8$ for mole fractions of entrapped liquid lower than $\sim$0.2--0.3. Above this range of values, C$_2$H$_6$ becomes again the most abundant volatile present in clathrate. On the other hand, the mole fractions of N$_2$, Ar and CO are in ranges close to those computed in the case of structure I clathrate formation. In both clathrate structures, the temperature plays the same role as the one noted for liquid reservoirs. At a similar species abundance, an increase of temperature decreases by several percent the corresponding mole fraction of entrapped liquid.

\section{Discussion and conclusion}

Because of the large uncertainties on the kinetics of clathrate formation (see Sec. 2.3), the reliability of our conceptual model requires future laboratory experiments at conditions close to those encountered on Titan in order to be assessed. If alkanofers equilibrated with clathrate layers, then our computations should allow disentangling in situ measurements of lakes and rivers flowing from alkanofers from those of liquid areas directly sourced by precipitation. In the case of structure I clathrate formation, and irrespective of the  temperature considered, the solution is dominated by ethane at mole fractions of the initial liquid reservoir trapped in clathrate lower than $\sim$0.9. At higher mole fractions, propane becomes the only species remaining in the liquid phase. In the case of structure II clathrate formation, ethane is the only dominant species in solution, irrespective of the temperature considered and mole fraction of entrapped liquid. These trends can be explained by the fact that ethane naturally enters the large cages of a structure I clathrate, while the large size of propane only allows this molecule to enter the large cages of a structure II clathrate. As both guests are very strong clathrate formers (e.g., Sloan \& Koh 2008), they compete for the same site. Therefore, in the case of formation of a structure I clathrate, propane remains in the liquid phase. Conversely, in the case of formation of a structure II clathrate, propane dominates in the large cages, forcing ethane to remain in the liquid phase. Our model then suggests that ethane and propane should be the discriminating markers of the clathrate structure forming from the solution when a significant fraction ($>$ 0.9) of the initial liquid reservoir has been entrapped. 

In our model, any river or lake emanating from alkanofers possessing these characteristics should present a similar composition. On the other hand, lakes and rivers sourced by precipitation should contain substantial fractions of CH$_4$ and N$_2$, as well as minor traces of Ar and CO (see Table \ref{lake0}). Here, the lake compositions have been computed in the cases of liquid trapping in structures I and II clathrate reservoirs. However, it has been shown that a mixture dominated by ethane in presence of methane leads to the formation of a structure I clathrate (Takeya et al. 2003). We then believe that clathrate reservoirs formed from the lakes on Titan should be essentially of structure I. 

Interestingly, if one postulates that the pores are fully filled by liquid hydrocarbons, it is possible to estimate the maximum porosity of the alkanofer that is consistent with a full enclathration of the solution. Assuming that the composition of the liquid reservoir is dominated by C$_2$H$_6$ in the case of structure I clathrate, the number of molecules present in the liquid is $N_L = \frac{\rho_{C_2H_6}~x V}{M_{C_2H_6}}$, with $x$ the porosity, $\rho_{C_2H_6}$ the density of the liquid, $M_{C_2H_6}$ the molecular mass of C$_2$H$_6$ and $V$ the volume of the alkanofer. On the other hand, the maximum number of available clathrate cages in the porous matrix is $N_c = \frac{\rho_{H2O}~(1-x) V}{7.66 M_{H_2O}}$, with $\rho_{H_2O}$ the density of solid water ice, $M_{H_2O}$ the molecular mass of H$_2$O, and 7.66 the number of H$_2$O molecules per enclathrated C$_2$H$_6$ molecule (Sloan \& Koh 2008). The value of $x$ that satisfies the conditions $N_L$/$N_c$ = 1 is the maximum porosity for which the number of trapped molecules is lower than or equal to the number of available clathrate cages. A larger value implies that all the liquid cannot be trapped as clathrates and a fraction of the liquid remains in the alkanofer. Assuming that the volume of ice remains constant during clathrate formation (Erfan-Niya et al. 2011), we find a maximum porosity value of $\sim$0.23. This value is larger than some estimates of 10\% to 15\% for the porosity of Titan's subsurface (Kossacki \& Lorenz 1996). The reservoir would be filled up until the ice is fully transformed into clathrates. Liquids filling the reservoir would then react with the clathrate matrix with exchange mechanisms such as those described in Choukroun \& Sotin (2012).
 
As mentioned in Sec. 2.3, our model can apply to the case where lakes and rivers located on Titan's surface directly equilibrate with a clathrate layer located beneath if the reequilibration timescale between the liquid and the atmosphere is short compared to the clathration timescale. Indeed, the massive  atmosphere would serve only to buffer methane and N$_2$ (and the minor species CO and Ar), but not ethane and propane. Because the vapor pressures of ethane and propane are so small, the atmosphere is not a reservoir of those species. Hence methane and other minor gaseous compounds would draw into the atmosphere when the ethane/propane go into the clathrate, and would be again introduced into the lake/sea when ethane/propane are added. So, the methane abundance in the seas would adjust so as to be in a thermodynamically correct proportion to the ethane and propane in the lakes/seas for the given methane atmospheric humidity. In these conditions, the abundances of volatiles trapped in clathrate would correspond to the values computed when the mole fraction of entrapped liquid is very low and the compositions of coexisting lakes/seas would be very close to those given in Table \ref{lake0} when the clathrate reservoir is absent.

In any case, it must be borne in mind that the possible range of initial compositions of the liquid in equilibrium with the atmosphere (the one used in the starting composition of our liquid reservoirs) remains poorly constrained at present. Mole fractions predictions of current lake models vary by tens of percents (e.g., Cordier et al. 2012; Tan et al. 2013; Glein \& Shock 2013). Our predictions of clathrate/liquid equilibrium compositions are valid in any case where the mole fraction of C$_2$H$_6$ is prominent in the initial liquid reservoir. For example, similar conclusions are found when using the lake composition determined by Tan et al. (2013) at 93.7 K as the starting one of the alkanofer ($\sim$54\% of C$_2$H$_6$, 32\% of CH$_4$, 7\% of C$_3$H$_8$, and 7\% of N$_2$). On the other hand, our results strongly vary when using the liquid composition calculated by Tan et al. (2013) at 90 K ($\sim$8\% of C$_2$H$_6$, 69\% of CH$_4$, 1\% of C$_3$H$_8$, and 22\% of N$_2$), which is very different from the one they obtained at higher temperature. In this case, irrespective of the clathrate structure, the solution is dominated by N$_2$ and CH$_4$ at high mole fractions of entrapped liquid. The results of our model are wrong since the predicted mole fraction of remaining N$_2$, which can be as high as 90\%, exceeds its solubility limit in hydrocarbons ($\sim$10\%; Hibbard \& Evans 1968). To derive the correct composition of the solution, we would need to compute the simultaneous equilibrium between the clathrate reservoir, liquid and the generated N$_2$--rich atmosphere. New experimental data obtained on the atmosphere--liquid equilibrium composition at Titan's conditions are needed to refine the expected composition of the lakes. Another limitation of the model, and of all models of the composition and stability of mixed clathrates on Titan's surface or subsurface, is related to the paucity of experimental data available to constrain both the Kihara potential parameters for many clathrate formers (e.g. Choukroun et al. 2013) and the fugacity of these gases at conditions relevant to Titan's surface.

\section*{Acknowledgements}

O. M. acknowledges support from CNES. M.C. and C.S. acknowledge support from the NASA Outer Planets Research Program. Part of this work has been conducted at the Jet Propulsion Laboratory, California Institute of Technology, under contract to NASA. Government sponsorship acknowledged.

\begin{table}[h]
\centering\caption{Chemical composition of lakes at the poles and the equator prior to clathration (adapted from Cordier et al. 2013)}
\begin{center}
\begin{tabular}{lcc}
\hline
 					&  Equator (93.65 K)					& Poles (90 K)			\\
\hline
C$_2$H$_6$   			&  $8.48\times 10^{-1}$  			& $8.03\times 10^{-1}$  		\\
C$_3$H$_8$   			&  $8.23\times 10^{-2}$  			& $7.79\times 10^{-2} $  		\\
CH$_4$   				&  $6.60\times 10^{-2}$  			& $1.13\times 10^{-1}$  		\\
N$_2$     				&  $3.82\times 10^{-3}$  			& $6.14\times 10^{-3}$ 		\\
Ar     				&  $3.67\times 10^{-6}$ 		& $5.74\times 10^{-6}$ 		\\
CO     				&  $2.68\times 10^{-7}$  			& $5.30\times 10^{-7}$  		\\
\hline 
\end{tabular}
\end{center}
\label{lake0}
\end{table}

\clearpage

\begin{table*}
\centering \caption{Antoine equation parameters}
\begin{tabular}{lllll}
\hline
Parameter   			& A			& B			& C			& Reference						\\
\hline
C$_2$H$_6$			& 4.50706		& 791.3		& -6.422		&	Carruth and Kobayashi (1973)		\\
C$_3$H$_8$			& 4.53678		& 1149.36		& 24.906		&	Helgeson and Sage (1967)		\\
CH$_4$				& 3.9895		& 443.028		& -0.49		&	Prydz and Goodwin (1972)		\\
N$_2$				& 3.7362		& 264.651		& -6.788		&	Edejer and Thodos (1967)			\\
Ar              			& 4.46903    	& 481.012    	& 22.156		&	McCain and Ziegler (1967)		\\
CO              			& 3.36514    	& 230.27   	& -13.14		&	Yaws and Yang (1989)			\\
\hline
\end{tabular}
\label{ABC}
\end{table*}

\clearpage

\begin{table*}
\centering \caption{Parameters for Kihara potentials}
\begin{tabular}{lllll}
\hline
Molecule   			& $\sigma_{K-W}$ (\AA)		& $\epsilon~_{K-W}/k_B$ (K)		& $a_{K-W}$ (\AA) 		& Reference						\\
\hline
C$_2$H$_6$			& 3.24693					& 188.181						& 0.5651				&	Sloan \& Koh (2008)				\\
C$_3$H$_8$			& 3.41670					& 192.855						& 0.6502				&	Sloan \& Koh (2008)				\\
CH$_4$				& 3.14393					& 155.593						& 0.3834				&	Sloan \& Koh (2008)				\\
N$_2$				& 3.13512					& 127.426						& 0.3526				&	Sloan \& Koh (2008)				\\
Ar              				& 2.9434     				& 170.50     					& 0.184 				&	Parrish \& Prausnitz (1972)		\\
CO              			& 3.1515     				& 133.61    					& 0.3976 				&	Mohammadi et al. (2005)			\\
\hline
\end{tabular}
\label{Kihara}
\end{table*}

\clearpage

\clearpage

\clearpage

\end{document}